\documentclass{PoS}

\title{The kinematics of S5 1803+784}

\ShortTitle{The kinematics of S5 1803+784}

\author{\speaker{N.A. Kudryavtseva},$^{1,2}$\thanks{Member of the International Max Planck Research
         School (IMPRS) for Radio and Infrared Astronomy at the Universities of Bonn and
         Cologne} ~S. Britzen$^1$, A.~Witzel$^1$, E.~Ros$^1$, M.F.~Aller$^3$, H.D.~Aller$^3$, R.M.~Campbell$^4$, J.A.~Zensus$^1$, A.~Eckart$^5$, J.~Roland$^6$, A.~Mehta$^7$\\
        \llap{$^1$} Max-Planck-Institut f\"ur Radioastronomie, Bonn, Germany\\
        \llap{$^2$} Astronomical Institute of St.-Petersburg State University, Saint-Petersburg, Russia\\
        \llap{$^3$} Astronomy Department, University of Michigan, USA\\
        \llap{$^4$} Joint Institute for VLBI in Europe, Dwingeloo, The Netherlands\\
        \llap{$^5$} I. Physikalisches Institut Universit\"at zu K\"oln, Cologne, Germany\\
        \llap{$^6$} Institut d'Astrophysique, Paris, France\\
        \llap{$^7$} International University Bremen, Bremen, Germany\\
        E-mail: \email{nadia@mpifr-bonn.mpg.de}}

\abstract{We present the results of a multi-frequency analysis of
the structural variability in the parsec-scale jet of the blazar
S5~1803+784. More than 90 epochs of observations at 6 frequencies
from 1.6 GHz up to 22 GHz have been combined and analyzed. We
discuss an alternative jet model for the source. In contrast to
previously discussed motion scenarios for S5~1803+784, we find
that the jet structure within 12 mas of the core can most easily
be described by seven ``oscillating'' jet features. We find that
the parameters of jet features, such as core separation, position
angle and flux density, change in a periodic way with a timescale
of about 4 years. We also find evidence for a correlation between
these parameters and the total flux density variability. We
suggest a scenario incorporating a periodic form of motion (e.g.
rotation, precession), with a non-negligible geometrical
contribution to explain the observational results.}

\FullConference{8th European VLBI Network Symposium\\
         September 26-29, 2006\\
         Toru\'n, Poland}

\begin{document}

\section{Introduction}
The BL Lac object S5~1803+784 is a compact flat-spectrum radio
source with z = 0.68 \cite{stickel_1993} and shows a misalignment
between kpc- and pc-scales\cite{britzen_2005a}. It has been
observed for more than 30 years at different frequencies and with
different resolutions (e.g.
\cite{eckart_1986,eckart_1987,schalinski_1988,witzel_1988,gabuzda_1992,kollgaard_1992,britzen_1995}).
These observations enable us to investigate the long-term
evolution of its flux and structure. The brightest jet component
in S5~1803+784 used to be one of the most prominent candidates for
a ``stationary'' component. However, as was shown in
\cite{britzen_2005}, the so-called stationary component ``moves''
or ``oscillates''.

\section{Observations}
We analyzed VLBI observations at six frequencies $\nu=$1.6, 2.3,
5, 8, 15 and 22 GHz (1.6 and 2.3 GHz: \cite{marcaide_1995} and
observations performed by us at 1.6 GHz; 5 GHz:
\cite{marcaide_1995,guirado_2001} and L.I.~Gurvits (priv. comm.);
8.4 GHz: \cite{perez_2000,ros_2000,ros_2001}; 15 GHz:
\cite{perez_2000,kellermann_1998,kellermann_2004}, the 2cm survey
webpage). We analyzed archival data and re-imaged all the epochs
and parameterized the flux density distribution at parsec scales
using gaussian functions which were fitted in the interferometric
visibilities. The fitting was performed using the {\it DIFMAP}
package \cite{difmap} for which we used circular Gaussian
components in order to avoid extended elliptical components and to
facilitate the component identification. In order to compare our
results with results described in the literature, we include the
model-fitting parameters collected from different papers
\cite{fey_1996,lobanov_2000,lister_2001,tateyama_2002,britzen_2005}.
In Fig~\ref{corr_c1} (bottom right) we present a list of all the
observational epochs we analyzed for the present work. The total
number of analyzed epochs is 94.

\section{Results}
\begin{figure}[b]
\includegraphics[width=7.5cm,clip]{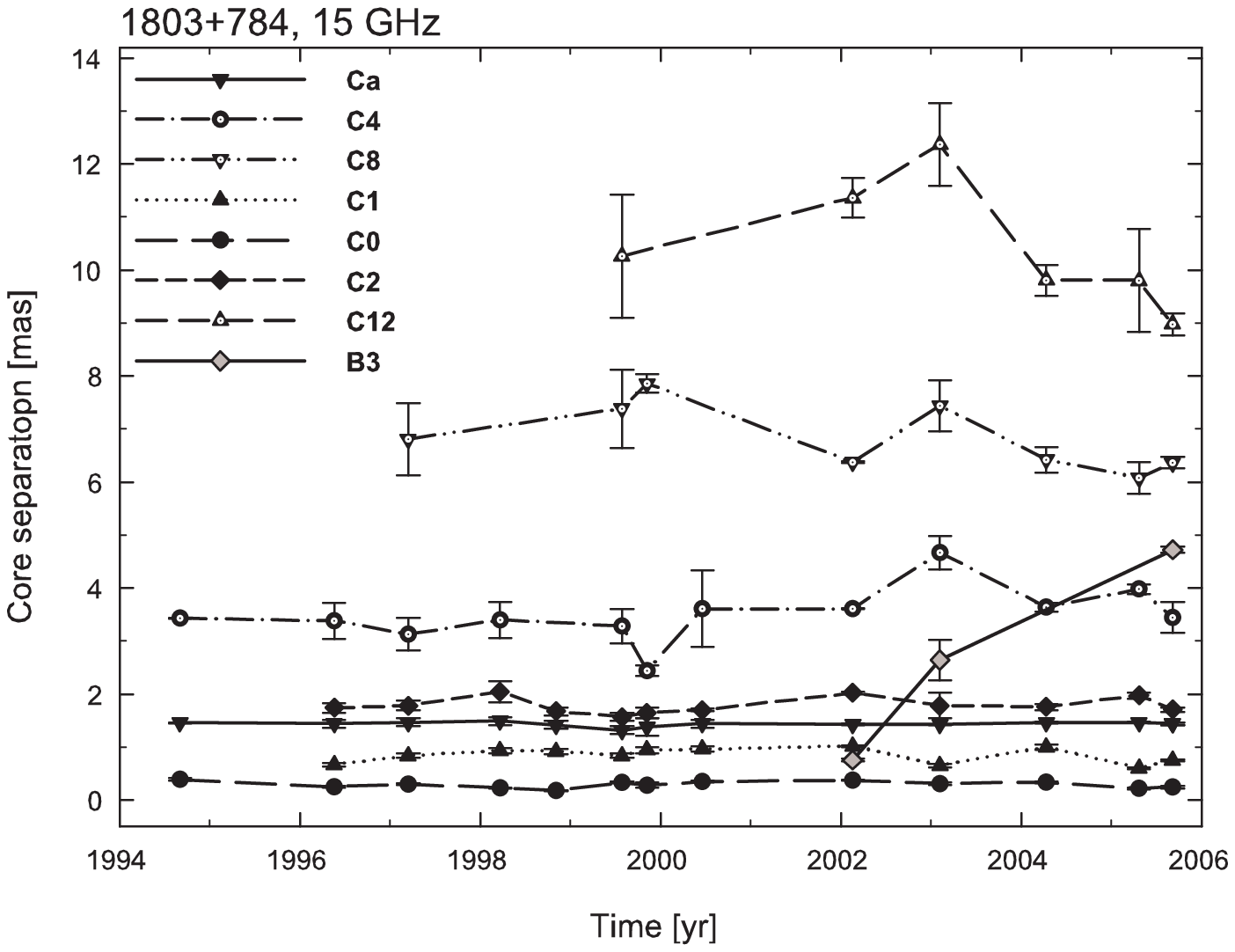}
\includegraphics[width=7.5cm,clip]{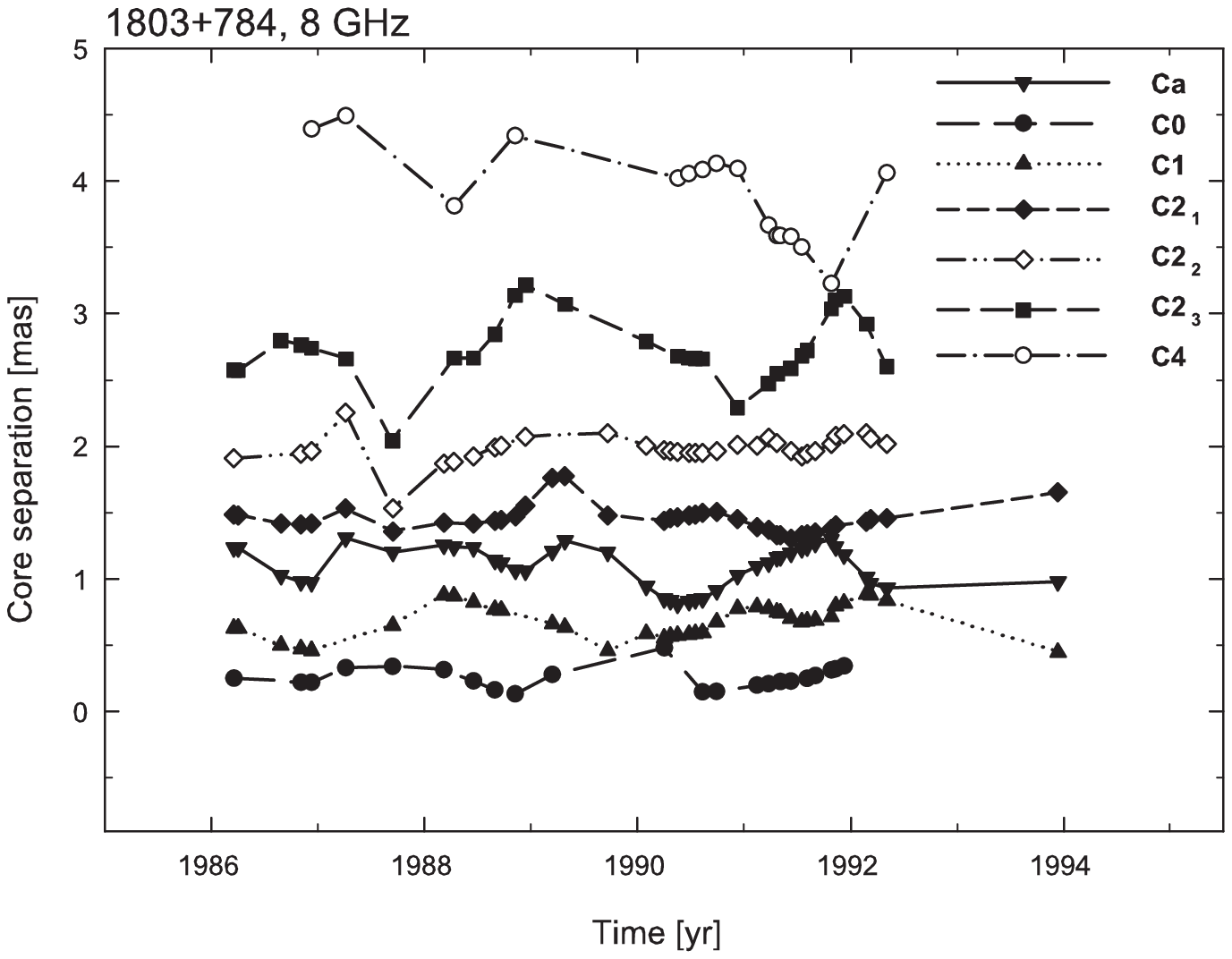}
\includegraphics[width=7.5cm,clip]{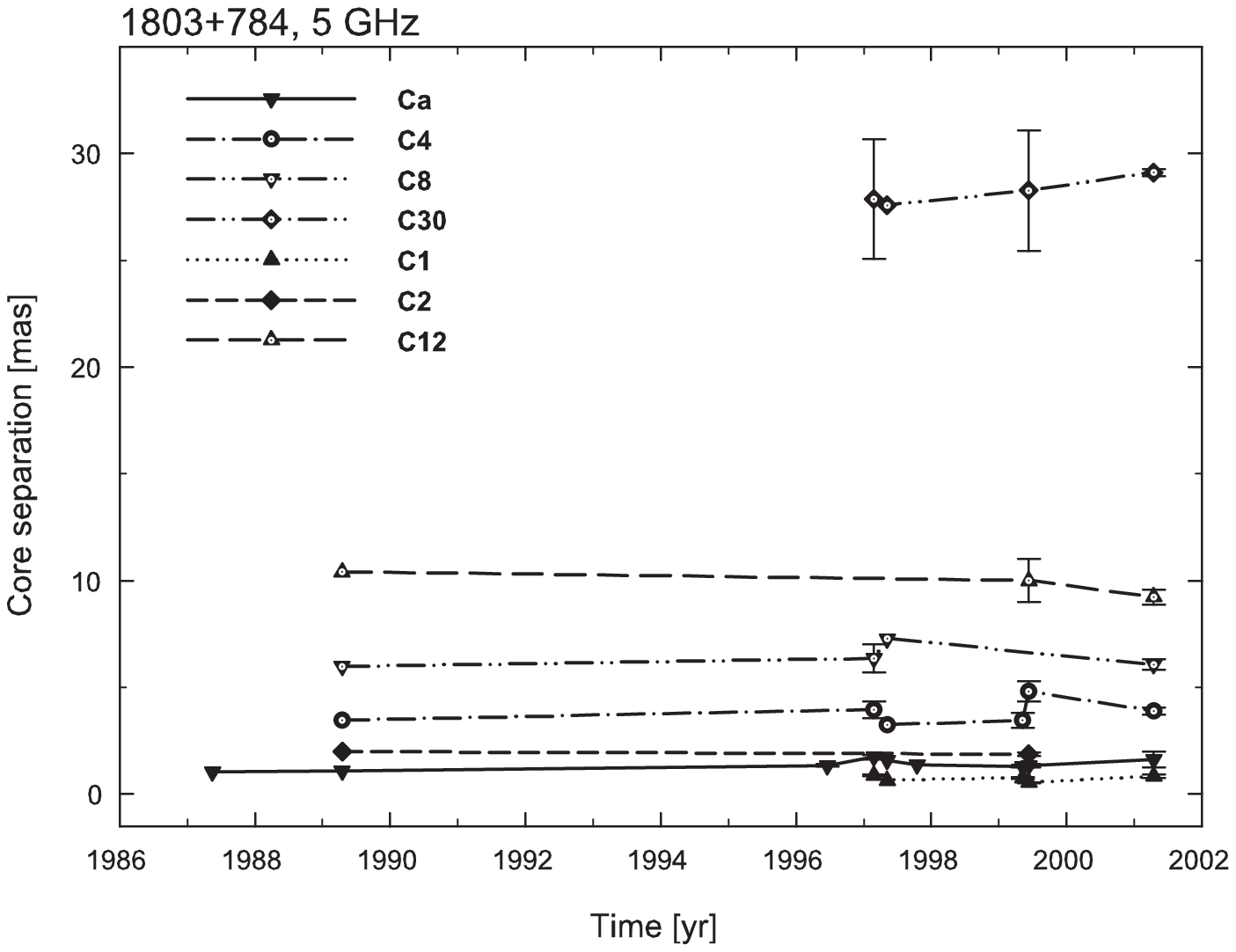}
\includegraphics[width=7.5cm,clip]{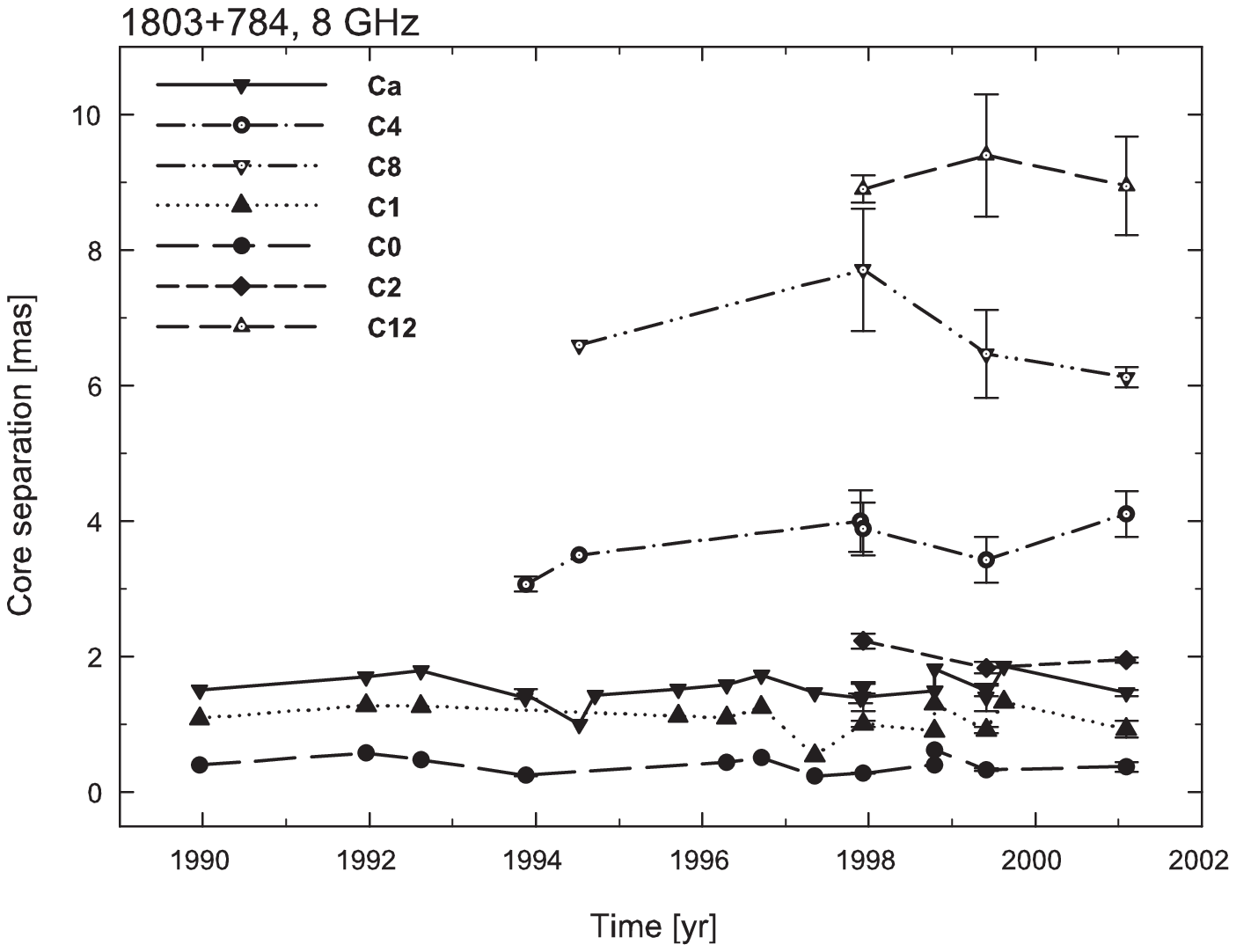}
\caption{\footnotesize We show the core separation as function of
time for those jet components detected at 15 GHz (upper left), 8
GHz (upper right for the period of time 1986 -- 1994; bottom right
for 1990 -- 2002) and 5 GHz (bottom left). Individual components
are denoted by different symbols and different lines.}
\label{identification}
\end{figure}

{\bf Oscillating components.} We performed a new identification
for the jet features of S5~1803+784, using the whole amount of
data. In Fig.~\ref{identification} we show the model-fitting
results at 5 GHz (bottom left panel), 8 GHz (top right for the
geodetic data from 1986 -- 1994; bottom right for data from the
literature) and 15 GHz (top left). It is clearly seen that in
addition to a ``stationary'' component {\bf Ca} 1.4 mas from the
core, there are several other components that also appear
stationary, both interior to {\bf Ca} ({\bf C0} at $\sim$ 0.3 mas,
{\bf C1} at $\sim$ 0.8 mas) and exterior to {\bf Ca} ({\bf C2} at
$\sim$ 2 mas, {\bf C4} at $\sim$ 3--4 mas, {\bf C8} at $\sim$ 6--8
mas, and {\bf C12} at $\sim$ 10--12 mas).

The jet components remain at similar core separations from 1986
until 2006, while the position angle of the components changes
with time (Britzen et al., in prep.). We find only one component
moving outwards during twenty years of observations, component
{\bf B3}, which is shown in the upper left in
Fig.~\ref{identification}. The jet of S5~1803+784 therefore
consists of a number of ``{\it oscillating}'' components, which do
not move outwards, but remain at an average value of the core
separation.

\begin{figure}[tb]
\includegraphics[width=10cm,clip]{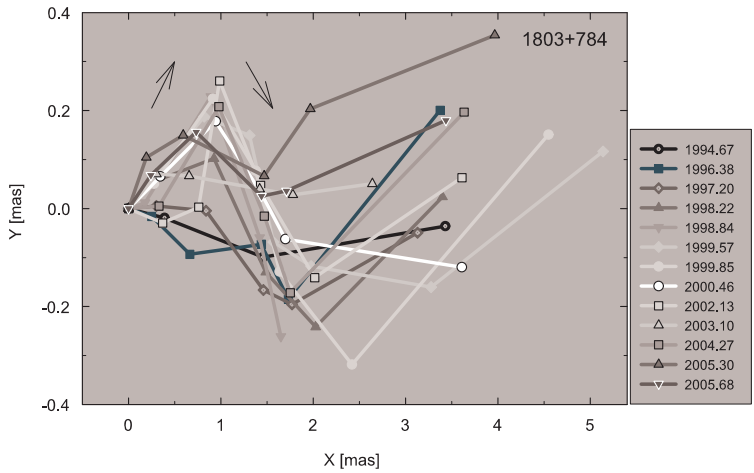}
\caption{\footnotesize The changes in the mean jet ridge line as a
function of time at 15 GHz are shown. Each dot represents the
position of one jet component in rectangular coordinates X and Y.
The lines connect all the components for one particular epoch of
observation. The shape of the jet changes from epoch to epoch from
an almost straight line to a sinusoidal form and than back to the
straight line again and thus possibly describes an oscillating
motion with a period of 8.5 years. The arrows indicate the
direction of changes. } \label{rotation}
\end{figure}

{\bf Quasi-periodicity.} The mean jet ridge line of S5~1803+784
changes gradually with time in a periodic way. Fig.~\ref{rotation}
shows the temporal evolution of jet component positions at 15 GHz
for the time period 1994 -- 2005. Each dot represents the position
of one jet component in rectangular coordinates X and Y. The lines
connect all the components for one particular epoch of
observation. From an almost straight line in 1994.67, the shape of
the jet evolves into a sinusoidal contour with a maximum positive
Y at X $\sim$ 1 mas and a maximum negative Y at X $\sim$ 2 mas.
The amplitude of the sinusoid reaches its maximal values in the Y
coordinate in 1998.84 decreases again, forming an almost straight
line in 2003.10. One period is completed after $\sim$ 8.5 years
and the jet shape starts to evolve into a sinusoid again. However,
the position of the straight lines in 1994.67 and 2003.10 are
different and the difference in the Y coordinate is 0.1 mas.

In order to check for possible quasi-periodic changes in the shape
of the jet, we applied a discrete autocorrelation function (DACF,
\cite{edelson_1988}) and the Jurkevich method
\cite{Jurkevich_1971} to the variations of the jet components'
parameters with time for the all available data from 1981 until
2005, including the core separation, the position angle and the
flux, from the inner-most component {\bf C0} to the outer-most
{\bf C12}. We find that a quasi-periodicity of about 4 years
exists in the variation of the core separation, the position angle
and the flux with time for the inner jet components {\bf C0}, {\bf
C1}, {\bf Ca} and {\bf C2} at 8 and 15 GHz. The DACF peak is at
the level 0.8--0.9. and the $f$ function which indicates the
trustworthiness of the identified periods\cite{kidger_1992}
($f>0.5$ indicates very strong periodicity) is from 0.51 to 1.80.
The 8.5-year cycle mentioned before was found for the time period
1994 -- 2005, whereas this 4-year period was found for the whole
period 1981 -- 2005. The 8.5-year cycle is twice the period found
by the time series analysis. A possible explanation of this fact
could be that the peak in 1999, in the middle of a 8.5-year cycle,
had an unusual low amplitude by comparison with other peaks. The
periodicity of 2 and 3.9 years was found before in total radio
flux density light curves \cite{kelly_2003} by the means of a
cross-wavelet transform and the second period is similar to a
four-year period found here.

\begin{figure}[ht]
\includegraphics[width=6.0cm,clip]{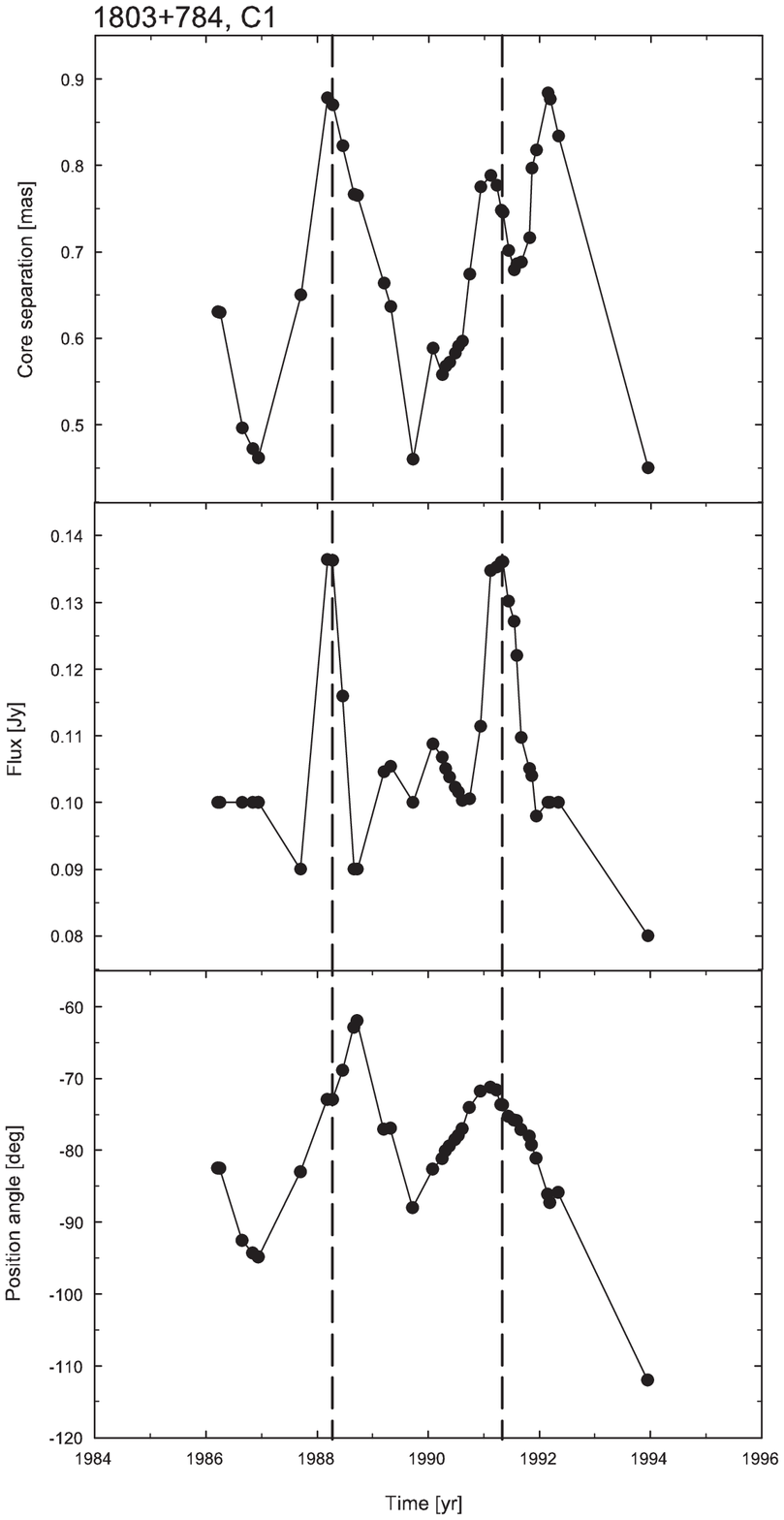}
\vbox{\hbox {\includegraphics[width=10.0cm,clip]{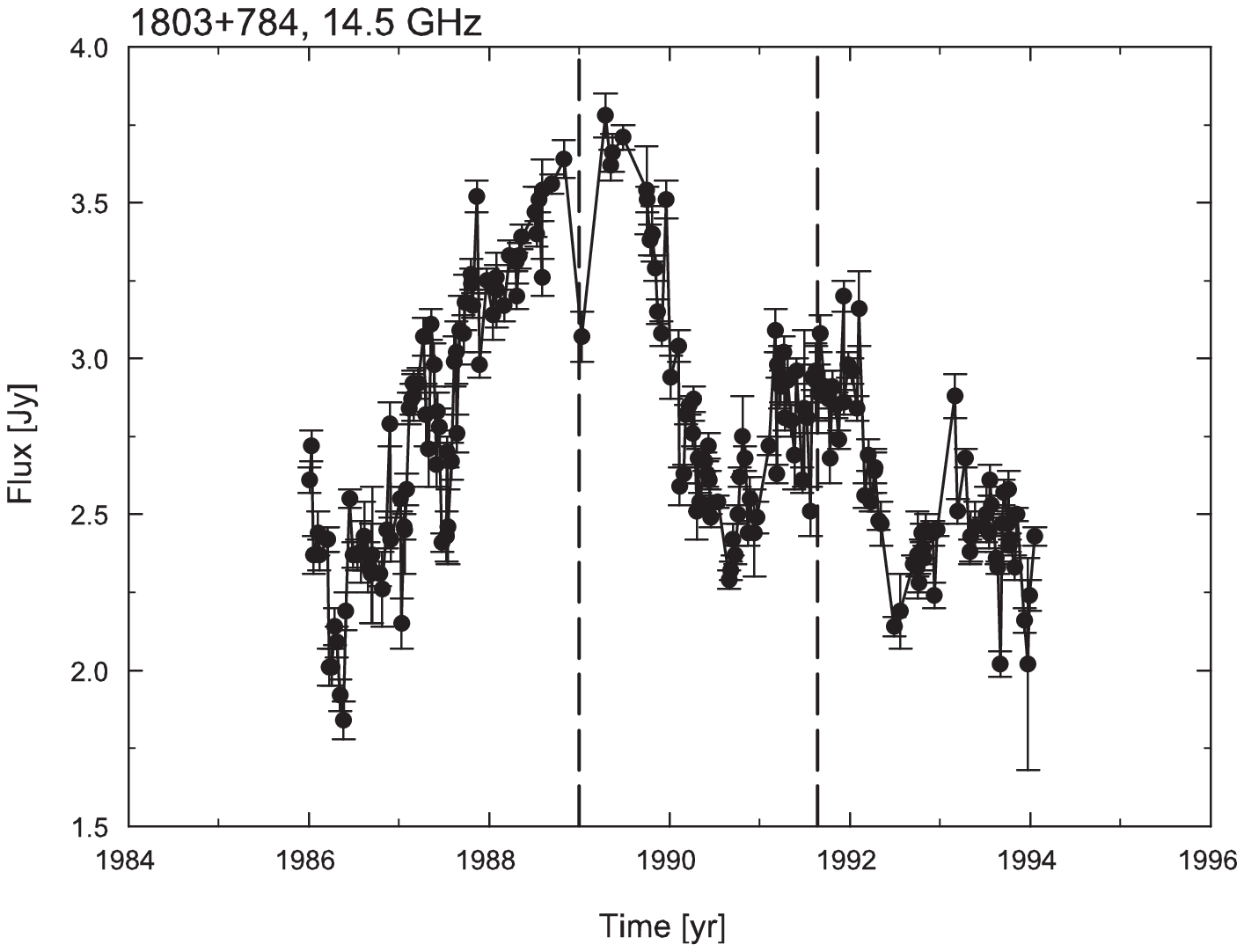}}
\hbox{
\includegraphics[width=9.0cm,clip]{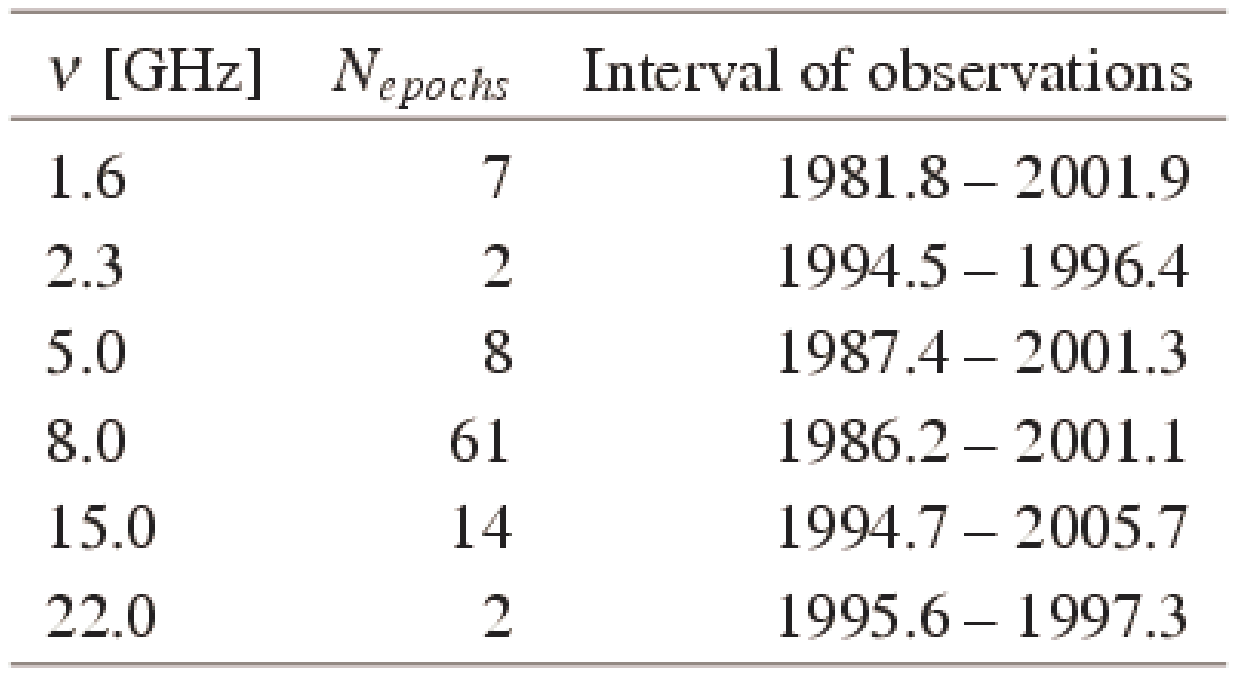}}}

\caption{\footnotesize {\bf Left:} The core separation, flux
density and position angle of a component {\bf C1} as a function
of time at 8 GHz during the period 1984--1996. It is clearly seen
that these three component parameters are correlated. Dashed lines
indicate the position of the peaks. {\bf Top Right}: Total flux
density light curve for 14.5 GHz (Aller et al. 1985). {\bf Bottom
Right}: The list of observational epochs of S5~1803+784, where
$\nu$ is the frequency of the observations in GHz, $N_{epochs}$ is
the number of epochs, and {\it Interval of observations} is the
time span covered by the observations.} \label{corr_c1}
\end{figure}

{\bf Correlations.} We determined the cross-correlation functions
\cite{edelson_1988} to search for possible correlations between
the variability of various jet parameters and the total flux
density. We calculated the cross-correlation functions for the
pairs of parameters within one of the components {\bf C0}, {\bf
C1}, {\bf Ca}, {\bf C2} and {\bf C4}, such as the core separation,
the flux density, the position angle changes and the total flux
density variability (UMRAO data, \cite{aller_1985}) at different
frequencies. For a few jet components such a correlation is
clearly visible. As an example, we show in Fig.~\ref{corr_c1}
(left) the variation of core separation, position angle and flux
density for the component {\bf C1} at 8 GHz in the period 1984 --
1996. We found that for all the inner components {\bf C0}, {\bf
C1}, {\bf Ca}, {\bf C2} and {\bf C4} at 8 and 15 GHz there is a
correlation between different parameters of components (peak of
discrete correlation function varies from 0.60 to 0.99 for
different parameters). At other frequencies, for which the data
are much more sparse and inhomogeneous, the correlation could not
be detected.

Moreover, a correlation exists not only between parameters of the
jet components, but also between the variation of the jet
parameters and the total flux density changes (see
Fig.~\ref{corr_c1}, top right). For the inner components {\bf C0},
{\bf C1}, {\bf Ca} and {\bf C2} the core separation and position
angle changes correlate with the total flux density light curves
at 8 and 15 GHz with correlation coefficients from 0.3 to 0.8 and
almost zero time delay. For component {\bf C4} we did not find any
correlation. A special case is component {\bf Ca}, which shows a
correlation between changes in the core separation and the total
flux density but an anti-correlation between changes in the
position angle and the total flux density behavior that applies to
8 and 15 GHz. Details of this analysis will be presented in
Britzen et al. (in preparation).

\section{Discussion}
We find that the jet can be described as a set of seven ``{\it
oscillating}'' features, which do not move outwards, but stay near
an average value of core separation. The jet component parameters,
such as core separation, position angle and flux density change in
a periodic way with a period of $\sim$4 years for the inner jet
components at 8 and 15 GHz. A similar 3.9-year period is found by
Kelly et al. (2003)~\cite{kelly_2003} in the total flux density
light curves. Moreover, changes of the core separation, position
angle and flux are correlated with each other and the total flux
density variability. Based on the results presented here, we
conclude that a scenario incorporating a periodic form of motion
(e.g. rotation, precession), with a non-negligible geometrical
contribution, can possibly describe the behavior of the jet
components. A detailed analysis of the kinematics of S5~1803+784
will be discussed in a forthcoming paper (Britzen et al., in
preparation).

\acknowledgments{ N.A. Kudryavtseva was supported for this
research through a stipend from the International Max Planck
Research School (IMPRS) for Radio and Infrared Astronomy at the
Universities of Bonn and Cologne. This work has benefited from
research funding from the European Community's sixth Framework
Programme under RadioNet R113CT 2003 5058187. UMRAO has been
supported by a series of grants from the NSF and by funds from the
University of Michigan. We are grateful to the group of the VLBA 2
cm Survey and the group of the MOJAVE project for providing the
data.}

\end{document}